\documentclass{epl}

\title{Tunneling Splittings in Mn$_{12}$-Acetate Single Crystals}
\author{E. del Barco\inst{1}, A. D. Kent\inst{1}, E. M. Rumberger\inst{2}, D. N. Hendrickson\inst{2}
\and G. Christou\inst{3}}
  \institute{
  \inst{1} Department of Physics, New York University - 4 Washington Place. New York, NY 10003\\
  \inst{2} Department of Chemistry and Biochemistry. University of California San Diego - La Jolla, CA 92093-0358\\
  \inst{3} Department of Chemistry, University of Florida - Gainsville, FL 32611-7200}
\pacs{75.45.+j,75.50.Tt}{Macroscopic quantum phenomena in magnetic
systems}

\begin{document}

\maketitle

\begin{abstract}
A Landau-Zener multi-crossing method has been used to investigate
the tunnel splittings in high quality Mn$_{12}$-acetate single
crystals in the pure quantum relaxation regime and for fields
applied parallel to the magnetic easy axis. With this method
several individual tunneling resonances have been studied over a
broad range of time scales. The relaxation is found to be
non-exponential and a distribution of tunnel splittings is
inferred from the data. The distributions suggest that the
inhomogeneity in the tunneling rates is due to disorder that
produces a non-zero mean value of the average transverse
anisotropy, such as in a solvent disorder model. Further, the
effect of intermolecular dipolar interaction on the magnetic
relaxation has been studied.
\end{abstract}

Single molecule magnets (SMMs) such as Mn$_{12}$, Fe$_8$ and
Mn$_4$ have enabled new studies of the physics of magnetic
nanostructures. These materials are ordered crystals of nominally
identical high spin ($S=$ 10 and smaller) molecules with
predominately uniaxial magnetic anisotropy. A particular focus has
been on quantum tunneling (QT) of the magnetization
\cite{1,2,3,4,5,6,7,8,9,10}, with efforts aimed at understanding
the origin of the tunneling phenomena in SMMs, the effect of
intermolecular interactions in SMM crystals and the role of the
environmental degrees of freedom in modulating QT (including
phonons and nuclear spins) \cite{11}. Even in the original, high
symmetry and high magnetic anisotropy barrier SMM,
Mn$_{12}$-acetate, the QT phenomena remains poorly understood.
Primarily this is due to the fact that the magnetic properties do
not reflect the tetragonal symmetry (S4 site symmetry) of the
molecule. For example, QT transitions do not obey the selection
rules imposed by this symmetry and the magnetic relaxation is
clearly non-exponential, indicating inhomogeneity in the tunneling
rates. It has become evident in recent research \cite{12,13} that
disorder that lowers the molecule symmetry plays an essential role
in the physics of QT in Mn$_{12}$-acetate single crystals, but the
precise nature of the disorder remains unclear. Thus far two
alternative models of disorder have been proposed. The first is
line-dislocations in crystals that produce long-range deformations
and through magnetoelastic interactions lead to a distribution of
the transverse magnetic anisotropy. This leads to a very broad
distribution in the tunnel splittings on a logarithmic scale
\cite{14}. The second is disorder in the solvent molecules that
surround each Mn$_{12}$ core that lead to a discrete set of
transverse anisotropies \cite{15}. Each model has predictions that
can be compared to experiment. In addition, intermolecular dipolar
interactions are another important source of magnetic
inhomogeneity that can be studied in experiment.

In this letter we present experimental results on the tunnel
splittings in deuterated Mn$_{12}$-acetate single crystals in the
pure quantum regime, in which relaxation is by QT without thermal
activation \cite{10}. Deuterated crystals have been studied
because the purity of the chemicals used in the synthesis leads to
very high quality crystals \cite{16}. We have used a modification
of the Landau-Zener (L-Z) method and a high sensitivity micro-Hall
effect magnetometer to study the magnetic relaxation in Mn$_{12}$.
With this method we have determined the distribution of tunnel
splittings for a \textit{single} QT resonance in a manner that
enables a critical comparison of the experimental data with models
of disorder for Mn$_{12}$. We also show the effect of
intermolecular dipolar interactions on the magnetic relaxation in
Mn$_{12}$.

The effective spin Hamiltonian for Mn$_{12}$ is
\begin{equation}
\label{e.1}{\cal {H}}=-DS_z^2-BS_z^4-g\mu _B{\bf {H}\cdot
{S}}+{\cal {H'}}\;\;,
\end{equation}
where the parameters $D=$ 0.548(3)K, and $B=$
1.17(2)$\times$10$^{-3}$K, have been determined by EPR \cite{7}
and inelastic neutron spectroscopy experiments \cite{8}. The first
two terms represent the uniaxial anisotropy of the molecules, the
third term is the Zeeman interaction of the spin with the magnetic
field, $\bf{H}$, and the fourth term includes dipolar and
hyperfine interactions and higher order transverse anisotropy
terms that do not commute with $S_z$. The observed steps in the
magnetic hysteresis correspond to an enhancement of the relaxation
towards the equilibrium magnetization when the eigenstates of
$S_z$, $m$ and $m'$, on opposite sides of the barrier are nearly
degenerate (see, for example, Fig. 1 of Ref. \cite{10}). These
resonances are labeled by the integer $k=m-m'$ and occur at
longitudinal magnetic fields of approximately $H_k=kD/(g\mu_B)\sim
k$ 0.42T.

It has been shown that the L-Z method is a powerful way to study
tunnel splittings in SMM crystals \cite{5}. The method consists of
sweeping the longitudinal magnetic field at a constant rate,
$\alpha=dH/dt$, across a QT resonance and measuring the fractional
change in magnetization. For an ensemble of non-interacting
identical SMMs, this fractional change in magnetization is related
to the probability that an individual SMM has \textit{remained} in
the original metastable state after an avoided level crossing. The
probability is $R=(M-M_{eq})/(M_{ini}-M_{eq})$, where $M_{ini}$ is
the magnetization before the crossing, $M_{eq}$ is the equilibrium
magnetization, and $M$ is the magnetization measured after
crossing the resonance. This probability is related to the quantum
splitting $\Delta$, through the L-Z formula
$R_{lz}=exp(-\pi\Delta^2/2\nu_0\alpha)$ , where
$\nu_0=g\mu_B(2S-k)$ and $\nu_0\alpha$ is the energy sweep rate.
Clearly, for this formula to be valid the energy sweep rate that
an individual SMM in a crystal experiences must be proportional to
the sweep rate of the applied external field. This will only be
true if changes in the internal magnetic field on each crossing
are small ($R\approx$ 1). For this reason, in our experiments, we
cross a resonance (say, the $k$-th resonance) $n$-times at a
constant and sufficiently fast sweep rate, so that on each
crossing $R\approx$ 1. Then we repeat this process for different
sweep rates. In this case, the probability \textit{to remain} in
the metastable state after the $n$-th crossing is $R_k=\frac
{M_n-M_{eq}}{M_{ini}-M_{eq}}$, where $M_n$ is the magnetization
after the $n$-th crossing of the resonance. The L-Z probability
after $n$-crossings is:
\begin{equation}
\label{e.2} {R_{lzn}=exp\left( -\frac {\pi\Delta^2}{2\nu_0}\frac
n\alpha\right)}\;\;,
\end{equation}
Based on this formula, relaxation curves recorded at different
sweeping rates should scale when plotted as a function of
$\alpha_{eff}=\alpha/n$. In figure 1 we show the data obtained for
resonance $k=$ 7 at different sweeping rates, $\alpha=$
3.33$\times$10$^{-3}$, 6.66$\times$10$^{-3}$ and
13.3$\times$10$^{-3}$T/s in the pure quantum tunneling regime for
this resonance, $T=$ 0.65K ($m=$ 10, $m'=$ -3). To do the
measurements we saturated the system with a negative high magnetic
field, $H=$ -5T, and then we swept the field to positive values
until we arrived to the vicinity of this resonance ($H_7=$ 3.55T),
with the applied field aligned with the $z$-axis of the crystal
\cite{17}. Then we crossed the resonance $n$ times ramping the
field up and down within $\pm$0.2 Tesla of the resonance at a
given rate. This procedure was then repeated for different sweep
rates. The scaling of the curves is clear, indicating that
$\alpha_{eff}$ is the relevant parameter to characterize the data
for this range of sweep rates. It is important to point out that
the relaxation curve does not follow eq.~(\ref{e.2}), i.e., it is
not an exponential function of $\alpha_{eff}$, which we discuss in
detail below.
\begin{figure}
\centering
\includegraphics[height=8cm]{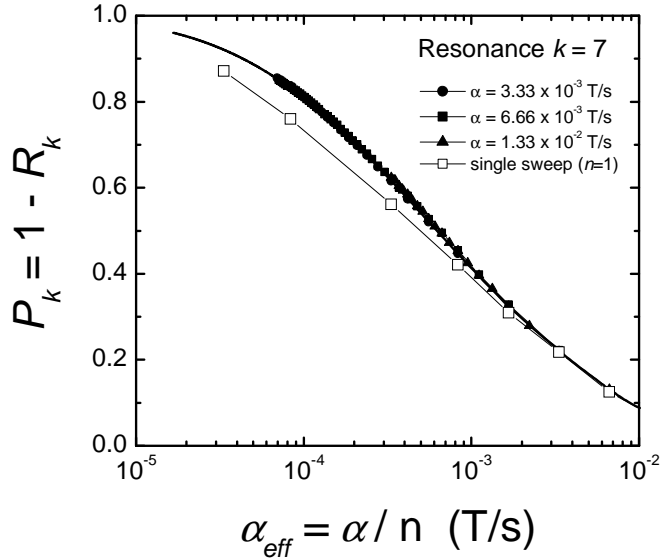}
\caption{The tunneling probability for deuterated Mn12-acetate as
a function of the effective sweeping rate, $\alpha_{eff}=\alpha/n$
of the applied magnetic field in a L-Z multi-crossing experiment
recorded at different sweeping rates (small closed symbols) for
resonance $k=$7 and at a temperature of 0.65 K. The open big
squares correspond to the tunnel probability in a L-Z
single-crossing experiment versus the sweeping rate ($n=$1). The
difference between the two curves we associate with dipolar
interaction between molecules \cite{19}.}\label{f.1}
\end{figure}

For smaller sweep rates, changing internal magnetic fields should
affect the magnetic relaxation. This was studied experimentally in
Fe$_8$ \cite{18}, where the relaxation departs from the L-Z
behavior for sweep rates below $\sim$10$^{-3}$T/s, and was
theoretically explained in terms of dipolar interactions by Liu et
al. \cite{19}. To study this in Mn$_{12}$, we have measured a
single-crossing L-Z relaxation at resonance $k=$ 7 for different
sweeping rates ranging from $\sim$10$^{-5}$ to $\sim$10$^{-2}$T/s.
The results are compared to the multi-crossing procedure in fig. 1
(open squares). The difference between these procedures is clear.
The break point from the L-Z single-crossing behavior is
$\alpha_c\sim10^{-3}$T/s and is similar to that observed in
Fe$_8$, while the deviations from the scaled data are smaller.
This is due to the fact that dipolar interactions in Mn$_{12}$ are
smaller than in Fe$_8$ because of differences in the
intermolecular distances and crystal structure \cite{20}. We
emphasize that the L-Z multi-crossing procedure permits studies at
very small effective sweep rates in a regime in which a single
crossing experiment would be complicated by intermolecular dipolar
interactions.
\begin{figure}
\centering
\includegraphics[height=8cm]{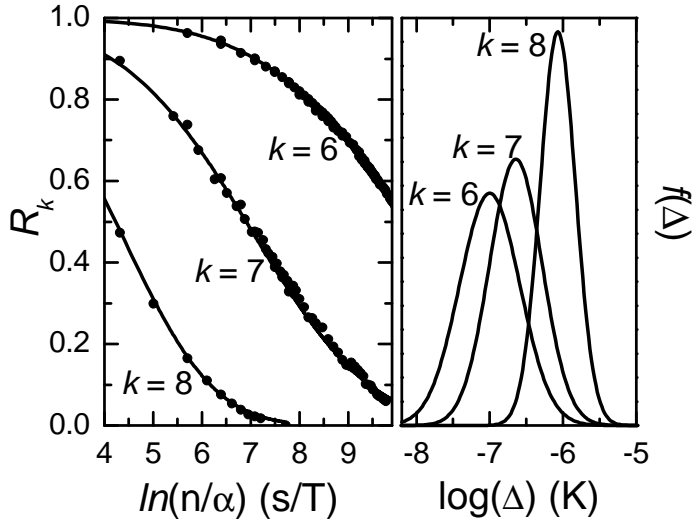}
\caption{(a) Probability to remain in the metastable well in a L-Z
multi-crossing experiment using different sweep rates
(3.33$\times10^{-3}$T/s to 1.33$\times10^{-2}$T/s) for different
resonances, $k$. Lines are fits using a log-normal distribution of
splittings. (b) Corresponding splitting distribution functions
extracted using the fitting procedure described in the text. The
centers, $x_c$, are -7.34, -6.54, and -6.07 and the widths, $W$,
are 0.43, 0.36, and 0.24, for resonances, $k=$6, 7, and 8,
respectively. }\label{f.2}
\end{figure}

We have carried out multi-crossings measurements of resonances
$k=$ 6, 7, and 8 at different sweep rates, $\alpha>\alpha_c$. The
results are presented in fig. 2a starting from a saturated sample.
Small differences were observed in the same procedure carried out
on three different single crystals synthesized in the same way.
The main results are the following: (a) The behavior of the
relaxation is clearly non-exponential. The relaxation is broad on
a logarithm scale, indicating the presence of a distribution of
quantum splittings in the crystal, and, (b) the data give direct
information on the distribution width due to the fact that
multi-crossings procedure allows us to measure a large fraction of
the relaxation curve for each resonance. We have assumed a
log-normal function to extract the distribution of the quantum
splittings in Mn$_{12}$. We take the form,
$f(x)=Aexp(-(x-x_c)/2W^2)$, with $x=log\Delta$, and we fit the
relaxation curves with the following expression,
\begin{equation}
\label{e.3} {R(\alpha,n)=\int_{-\infty}^\infty
R_{lz}(\alpha,\Delta,n)f(log\Delta)dlog\Delta}\;\;,
\end{equation}
The fitting parameters are the center of the distribution $x_c$
and the width of the distribution on a log scale, $W$. The fits
with eq.~(\ref{e.3}) are represented by lines in fig. 1 and fig.
2a and are in excellent accord with the experimental data. The
fitting parameters can be determined within an accuracy of 5$\%$.
The corresponding splitting distribution functions for each
resonance are shown in fig. 2b. The mean value of the distribution
increases with $k$ while the width remains more or less constant,
being somewhat narrower for $k=$ 8. This width is a factor of 2
(in $log\Delta$) smaller than the width extracted in
ref.\cite{12}, because of the higher quality of the deuterated
crystals.
\begin{figure}
\centering
\includegraphics[height=8cm]{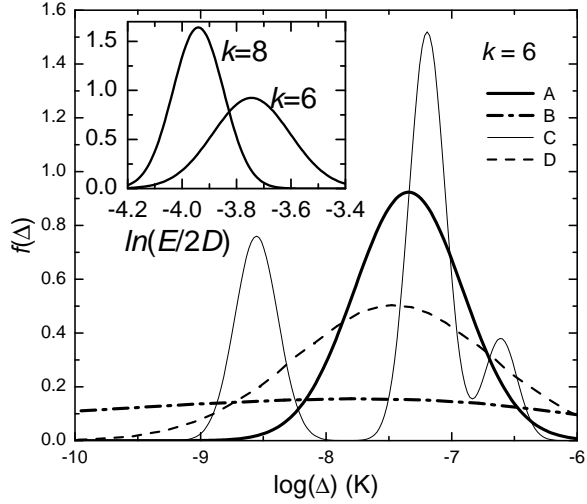}
\caption{Splitting distributions function extracted using (A) a
log-normal distribution, (B) the dislocation model, and (C) the
solvent disorder model. Distribution (D) was measured in a
non-Deuterated Mn$_{12}$-acetate single crystal (from
Ref.\cite{12}). The inset shows the distribution of the second
order anisotropy parameter, $E$, extracted from the log-normal
distribution of tunneling splittings using eq.~(\ref{e.5}) for $k$
= 6 and $k$ = 8.}\label{f.3}
\end{figure}

Recently it has been proposed that line dislocations \cite{14} or
solvent disorder \cite{15} should introduce a second order
anisotropy term, $E(S_x^2-S_y^2)$ in the Hamiltonian of Mn$_{12}$,
that increases the quantum tunneling probability. Line
dislocations generate a broad distribution of the parameter $E$
with mean value $<E>$ = 0, while disorder of the solvent molecules
creates a discrete number of $E$ values due to different Mn$_{12}$
isomers existing in the crystal. Recent relaxation experiments
have confirmed the existence of a broad splitting distribution
\cite{12,13} as well as broad absorption lines \cite{21} and
discrete double absorption peaks \cite{15,22}have been observed by
EPR experiments. In order to compare our experimental data to
these models we have tried to fit our relaxation curves to the
distributions predicted by these models. In the distribution
function predicted by the line dislocations model the mean value
and width of the distribution are not independent variables,
\begin{equation}
\label{e.4} {f_L(x)\cong \frac
{1}{2\sqrt{\pi}\tilde{E}_{\tilde{c}}} exp\left(x-\frac {e^{2x}}{(2
{\tilde{E}_{\tilde{c}}})^2}\right)}\;\;,
\end{equation}
where $x\equiv\ln\tilde{E}$ with $\tilde{E}=E/2D$.
$\tilde{E}_{\tilde{c}}$ is the width of the distribution of the
anisotropy parameter $\tilde{E}$. $\tilde{E}_{\tilde{c}}$ depends
on the geometry of the crystal and on the concentration of
dislocations per unit cell, $c$. To fit to this function we use
the following relation between $\Delta$ with $E$ which follows
from perturbation theory,
\begin{equation}
\label{e.5} {\ln\left(\Delta_k/g_k\right)/\xi_k=\ln\left(\frac
E{2D}\right)}\;\;,
\end{equation}
where $g_k$ and $\xi_k$ are determined by $k$, $S$, and $D$
\cite{14}. The only fitting parameter is $\tilde{E}_{\tilde{c}}$.
The result of this fit for resonance $k=$ 6 is shown in fig. 3
(dashed-dot line). It corresponds to a concentration of
dislocations of around $c\sim10^{-4}$. The mean value is chosen to
coincide with the center of the log-normal distribution function
(continuous thick line). Clearly, the width is many orders of
magnitude too large to fit the experimental data. Physically this
is due to the fact that the transverse anisotropy distribution
expected from the dislocations model alone is centered at $<E>$ =
0, which produces a large tail to small values of the tunnel
splitting, and would produce a much broader relaxation curve than
observed in experiment. As a consequence it is clear that a model
that describes our data must have non-zero mean value of the
transverse anisotropy distribution. For example, taking our
log-normal distribution of $\Delta$, the distribution of the
transverse anisotropy parameter can be inferred. The results for
$k$ = 6 and $k$ = 8 are shown in the inset of figure 3. As these
curves do not overlap (scale) it is clear that the origin of the
tunnel splittings cannot $\it{only}$ be due to a second-order
transverse anisotropy. However, qualitatively, this would lead to
a distribution of $\Delta$ that narrows for increasing $k$ (which
can be seen from perturbation theory, eq.~(\ref{e.5})), as
observed in fig. 2.

The solvent disorder model suggests a multi-peak discrete
splitting distribution due to the presence of six different
isomers of Mn$_{12}$ with different $E$ parameters. We have found
that it is not possible to fit the experimental data with a
sequence of delta-functions for the suggested values of $E$ for
each isomer. However, we have obtained a good fit to the data
assuming a log-normal distribution with a very narrow width for
each delta-function proposed in ref.\cite{15}. The resulting
distribution function has three well defined peaks as can be
observed in fig. 3 (continuous thin line). The values of the peak
centers are $x_{c,1}=$-7.19(-7.0525), $x_{c,2}=$-8.55(-8.1749),
and $x_{c,3}=$-6.60(-6.7995), in accord with the reported values
between parenthesis \cite{23}. The width of each peak is
$W_i=x_{c,i}$/50. The height of each peak is taken to be
proportional to the measured population of its corresponding
isomer. The data suggest that the solvent disorder model, which
gives a non zero value of the transverse anisotropy, together with
some other type of disorder (like line dislocations or point
defects) that introduces a small broadening of the anisotropy E
parameter around these values are consistent with our relaxation
data. We note other models that include higher order transverse
anisotropies (such as $C(S_x^2-S_y^2)^2$) and disorder of some
type will likely also have features necessary to understanding the
relaxation in Mn$_{12}$. Of course, in order to explain the
relaxation for odd resonances there must be transverse fields due
to small tilts of the anisotropy axis of the molecules or slight
misalignments of the applied field.

To conclude, we have used a powerful method to study quantum
splittings of SMM single crystals. We have reported for the first
time the break-down of L-Z scaling due to dipolar interactions in
Mn$_{12}$-acetate and we have estimated the order of magnitude of
the interaction energy between molecules. Our results suggest that
the quantum splitting distribution in Mn$_{12}$-acetate should be
due to disorder which generates a transverse anisotropy
distribution with a non-zero mean value, such as solvent disorder.

\acknowledgments The authors grateful acknowledge useful
discussions with S. Hill, P. Stamp and J. M. Hernandez. This
research was supported by NSF-NIRT Grant No. DMR-0103290 and
NSF-IMR-0114142. E. del B. acknowledges support from fellowship
provided by S.E.E.U. of Spain and Fondo Social Europeo.


\begin{thebibliography}{0}
\bibitem{17} The $c$-axis of the sample was aligned with an
accuracy of $\pm$3$^o$ by positioning crystals on a
Hall-magnetometer with the aid of an optical microscope. We note
that even such small missalignments could have an effect on the
tunneling rates, particularly for odd resonances.
\bibitem{23} The splitting values corresponding to two of the isomers of
Mn$_{12}$ are almost identical \cite{15}. We have included them as
only one peak with a height proportional to the sum of the
populations of these isomers. We note that this model also
includes a small fourth order transverse anisotropy term.
\end{thebibliography}
\end{document}